\documentclass[pdflatex,sn-chicago]{sn-jnl}

\usepackage{amsmath}
\usepackage{graphicx}
\usepackage{microtype}
\usepackage{xcolor}
\usepackage{manyfoot}
\usepackage{float}

\begin{document}

\title[The Algorithmic Blind Spot]{The Algorithmic Blind Spot: Bias, Moral Status, and the Future of Robot Rights}

\author*[1]{\fnm{Rahulrajan} \sur{Karthikeyan}}\email{rkarthi5@asu.edu}
\equalcont{These authors contributed equally to this work.}

\author[2]{\fnm{Moses} \sur{Boudourides}}\email{Moses.Boudourides@northwestern.edu}
\equalcont{These authors contributed equally to this work.}

\affil*[1]{\orgdiv{Computer Engineering Graduate Program},
  \orgname{Arizona State University},
  \orgaddress{\city{Tempe}, \postcode{85287-8809}, \state{AZ}, \country{USA}}}

\affil[2]{\orgdiv{School of Professional Studies},
  \orgname{Northwestern University},
  \orgaddress{\city{Evanston}, \postcode{60201}, \state{IL}, \country{USA}}}

\abstract{Contemporary debates in AI ethics increasingly foreground the prospective moral status of artificial intelligence and the possibility of extending moral or legal rights to artificial agents. While such discussions raise substantive philosophical questions, they often proceed alongside a comparatively limited engagement with the empirically documented harms generated by algorithmic systems already embedded within social, legal, and economic institutions.  
We conceptualize this asymmetry as an algorithmic blind spot: a discursive-structural pattern in which disproportionate ethical investment in speculative future artificial agents marginalizes empirically documented and asymmetrically distributed harms affecting human populations.
The paper analyzes prominent strands of the robot rights literature and juxtaposes them with empirical evidence of algorithmic bias and harm across domains including employment, criminal justice, surveillance, and facial recognition. It demonstrates how ethical preoccupation with hypothetical future entities can obscure existing injustices, diffuse responsibility, and impede mechanisms of accountability and redress. Without rejecting philosophical inquiry into the moral status of artificial systems, the paper instead emphasizes the importance of ethical prioritization and temporal ordering within AI ethics. Addressing the algorithmic blind spot, we argue, requires re-centering ethical evaluation on human impacts, institutional responsibility, and the governance of algorithmic systems currently in operation. In doing so, the paper introduces a conceptual framework for critically assessing ethical discourse in AI and underscores the need to align ethical reflection more closely with its immediate social consequences.}


\maketitle

\section{Introduction: The Divided House of AI Ethics}\label{sec1}

Contemporary debates in artificial intelligence (AI) ethics are shaped by a persistent and increasingly consequential internal tension. One major strand of discussion is oriented toward speculative philosophical inquiry into the possible future moral status of artificial systems. Alongside this, a parallel body of scholarship and policy engagement confronts the immediate and empirically documented harms produced by AI systems already deployed across social, economic, and political domains. Although both strands raise legitimate ethical concerns, differences in their framing and prominence have generated a systematic imbalance in moral attention.

The speculative strand of AI ethics focuses primarily on foundational questions concerning moral agency, moral patienthood, and the conditions under which artificial entities might warrant moral consideration or legal rights. These debates draw extensively on philosophical discussions of consciousness, sentience, personhood, and moral status, as well as on thought experiments involving artificial general intelligence and superintelligent systems \citep{Floridi2013,Danaher2020}. Their intellectual lineage extends back to early reflections on machine intelligence, most notably \citet{Turing1950}'s question of whether machines can think, and they have gained renewed momentum through rapid advances in machine learning and widespread cultural fascination with AI futures. Through philosophical argument, speculative ethics, and science-fiction-inspired imaginaries, this literature explores the possibility that artificial systems might one day qualify as moral patients or even moral agents.

These debates are both philosophically sophisticated and culturally influential. They shape public narratives, inform policy speculation, and increasingly influence academic and media discourse surrounding AI governance \citep{Cave2019}. At the same time, they are oriented by design toward hypothetical futures. The entities at the center of these discussions, including sentient machines, conscious algorithms, and autonomous moral agents, remain speculative at present. As a result, ethical urgency is grounded in anticipation rather than in observed social effects.

By contrast, a second and more immediately consequential strand of AI ethics addresses the widespread phenomenon of algorithmic bias and discrimination. This body of scholarship documents how AI systems used in hiring, credit scoring, policing, healthcare, and welfare administration reproduce and amplify existing social inequalities \citep{ONeil2016,Eubanks2018,Benjamin2019}. Empirical investigations have identified gender bias in automated recruitment tools \citep{Dastin2018,Andrews2022}, racial disparities in facial recognition technologies \citep{Buolamwini2018}, and systemic inequities embedded in criminal risk-assessment algorithms \citep{Angwin2016}. These harms are neither speculative nor evenly distributed. They occur in the present and disproportionately affect marginalized and historically disadvantaged populations.

The disproportionate ethical emphasis placed on speculative questions about future artificial moral agents contributes to what can be described as an \emph{algorithmic blind spot}. This blind spot refers to a structured pattern within AI ethics discourse in which moral concern is redirected toward hypothetical artificial subjects, while the concrete and ongoing harms imposed on human populations by existing algorithmic systems receive comparatively less sustained attention. It is not merely an oversight. Rather, it reflects a deeper normative and epistemic asymmetry in how ethical salience is assigned, privileging imagined futures over documented present realities.

The underlying problem does not lie in the philosophical legitimacy of inquiries into machine moral status. Instead, it arises from the consequences of sustained fascination with hypothetical artificial moral patients. Speculative narratives and abstract philosophical reasoning exert material influence over research agendas, funding priorities, and regulatory focus \citep{Whittaker2018,Baker2022}. A paradox emerges in which extensive ethical concern is devoted to the potential suffering of future machines, while the actual suffering of humans subjected to biased, opaque, and unaccountable algorithmic systems remains insufficiently addressed. This imbalance reflects a failure of moral prioritization rather than a failure of moral imagination.

The divergence between these strands of AI ethics also reflects deeper theoretical differences. Robot rights debates typically draw on liberal moral and political philosophy, emphasizing individual moral subjects, intrinsic properties, and rights-bearing status. Scholarship on algorithmic bias, by contrast, foregrounds structural injustice, institutional responsibility, and the collective dimensions of harm. The focus is less on the emergence of new moral subjects and more on ethical failures within existing socio-technical systems \citep{Benjamin2019,Wang2024}. These orientations carry distinct political implications. While debates about artificial moral status often remain speculative and academic, concerns about algorithmic bias have already prompted legal challenges, regulatory interventions, and public contestation.

Correcting this imbalance requires a reorientation of AI ethics toward a framework that places human well-being, dignity, and justice at its normative core. Such a reorientation does not foreclose future consideration of artificial moral status, but it insists that ethical inquiry remain anchored in present realities and grounded in responsibility for existing harms. Ethical deliberation cannot credibly prioritize the rights of hypothetical machines while deployed systems actively undermine human rights and democratic accountability. Recognizing and addressing the algorithmic blind spot is therefore a necessary condition for a responsible and socially responsive approach to AI ethics.

The argument unfolds in four stages. Section 2 develops the concept of the algorithmic blind spot and situates it within relevant theoretical traditions. Section 3 examines documented forms of algorithmic bias and their human consequences. Section 4 synthesizes these strands by operationalizing and empirically assessing the blind spot through bibliometric analysis, while addressing key counterarguments. Section 5 outlines a human-centered framework and derives institutional implications.

\section{The Allure of the Artificial Moral Patient: A Critical Review of the Robot Rights Debate}\label{sec2}

Before examining the principal positions within the robot rights debate, it is necessary to introduce the concept of the algorithmic blind spot, which provides a central analytical lens. The algorithmic blind spot is a discursive-structural pattern in AI ethics in which disproportionate ethical investment in speculative future artificial agents systematically marginalizes and de-prioritizes empirically documented and asymmetrically distributed harms affecting human populations. This pattern does not arise from individual intention, but operates through entrenched epistemic, institutional, and power-laden frameworks that render present algorithmic injustices comparatively less urgent, less visible, and less amenable to regulatory intervention. The concept is analytic rather than accusatory: it describes a structural configuration of ethical discourse, not the intentions or moral commitments of individual scholars.

The blind spot can be said to operate when three conditions converge. First, speculative debates concerning artificial moral status achieve comparable or greater discursive prominence than empirically grounded analyses of algorithmic harm. Second, documented human harms are framed primarily as technical imperfections rather than as structural injustices requiring normative prioritization. Third, ethical discourse assigns anticipatory urgency to hypothetical machine suffering while treating present, asymmetrically distributed human harms as secondary, deferrable, or as externalities of technological innovation. These criteria do not assume zero-sum displacement, but identify a patterned misalignment in ethical salience and prioritization.  

At this stage, the concept is introduced descriptively; its broader normative implications will be developed later in the analysis.

This concept aligns closely with established frameworks in Science and Technology Studies and political theory. One important point of convergence appears in Latour's critique of modern abstraction, particularly his argument that technological artifacts should not be understood as neutral tools but rather as elements embedded within heterogeneous networks of humans, institutions, values, and material practices \citep{Latour1993,Latour2005}. Ethical approaches that treat AI systems as isolated candidates for moral evaluation, assessed primarily through intrinsic properties such as consciousness or agency, reproduce what Latour characterizes as a purification process. This process separates technical artifacts from the social relations and power structures that give them meaning and effect. 

The concept also resonates with Foucauldian analyses of power, knowledge, and problematization. Foucault's work emphasizes how regimes of knowledge shape what is recognized as a legitimate object of ethical and political concern while rendering other forms of harm invisible or normalized \citep{Foucault1977,Foucault1980}.  
When AI ethics is framed primarily around speculative future subjects, ethical inquiry is subtly redirected away from existing regimes of algorithmic governance that discipline populations, structure access to opportunities, and redistribute risk. Issues such as bias, surveillance, and exclusion are consequently reframed as secondary technical effects rather than as central political and ethical problems.

A further theoretical connection emerges through Sylvia Wynter's critique of the overrepresentation of ``Man'' as the normative subject of moral and political theory \citep{Wynter2003}. Wynter argues that modern humanism universalizes a historically specific figure that is Western, liberal, and bourgeois, while marginalizing other forms of human life. Within AI ethics, debates over robot rights risk producing a dual abstraction. Moral concern is displaced onto speculative artificial entities, while the human populations most affected by algorithmic systems are often those already excluded from dominant conceptions of the human. 

The concept is also compatible with insights from post-normal science, which emphasizes that ethical reasoning in contexts marked by high uncertainty, high stakes, and value disagreement must remain attentive to social consequences, power asymmetries, and the distribution of risk \citep{Funtowicz1993}. Robot rights debates frequently operate under conditions of profound uncertainty regarding future machine capacities. Post-normal science cautions that under such conditions, ethical attention should not be monopolized by speculative scenarios at the expense of empirically grounded harms.

The algorithmic blind spot reflects a failure to apply these sensibilities consistently. Uncertainty about future AI consciousness is treated as morally urgent, while uncertainty borne by humans subjected to opaque and consequential algorithmic systems is treated as acceptable or unavoidable. 
It arises from technical framings of AI as autonomous and bounded agents evaluated through individualistic criteria, and it mirrors the operational logic of many deployed algorithmic systems that optimize narrowly defined objectives while externalizing social context and downstream harm. Ethical discourse that privileges speculative artificial moral patients over situated human impacts risks reproducing, at the normative level, the same abstractions and exclusions that characterize problematic algorithmic decision-making in practice \citep{ONeil2016,Eubanks2018,Benjamin2019}.

Recognizing the algorithmic blind spot does not entail rejecting inquiries into machine moral status altogether. Rather, it identifies a problem of moral prioritization and temporal inversion in which ethical attention is disproportionately allocated to uncertain future entities while present and well-documented harms to humans, particularly those affecting marginalized populations, are treated as secondary or merely technical. Within this framing, the robot rights debate can be read not only as a philosophical inquiry into moral status, but also as a site where this blind spot becomes visible. The discussion that follows examines how property-based theories, precautionary arguments, and even some relational approaches may unintentionally contribute to this displacement of ethical priority, while also highlighting conceptual resources capable of re-centering human vulnerability, institutional responsibility, and socio-technical power relations.

The question of whether machines could or should be granted rights remains one of the most compelling and contested issues in contemporary AI ethics. It stretches the boundaries of moral imagination by challenging deeply held assumptions about personhood, moral agency, and justice \citep{Floridi2013,Bryson2018}. At the same time, the robot rights debate does not constitute a single unified position. It encompasses a heterogeneous set of philosophical traditions and an expanding body of empirical research. Central arguments are accompanied by skeptical and social-relational responses that have emerged in reaction to more speculative claims. Despite its philosophical sophistication and cultural visibility, robot-rights scholarship often adopts a forward-looking orientation and emphasizes the intrinsic properties of artificial systems. This orientation can divert attention away from immediate ethical risks associated with real-world AI deployment. The stakes therefore extend beyond abstract theory, as public narratives, policy debates, and even system design priorities are shaped by how robot rights are framed \citep{Cave2019,Danaher2020}. Careful and sustained critique is thus required, one that remains attentive to both conceptual nuance and social consequence.

\section{The Present Danger: Algorithmic Bias and Its Human Cost}\label{sec3}

While discussions of robot rights often take place within abstract philosophical debates or futurist speculation, algorithmic bias represents an immediate and demonstrable threat. It is a concrete and empirically documented harm that already shapes the lives of millions of people worldwide. Biased AI systems are not a distant possibility but a present reality, actively deployed across high-stakes domains such as criminal justice, employment, healthcare, finance, and security \citep{ONeil2016,Eubanks2018,Benjamin2019}. Algorithmic bias can be examined through an analysis of its underlying sources, its real-world consequences, and its theoretical foundations. The harms produced by biased systems are not only more immediate and widespread than the speculative risks emphasized in robot-rights debates, but also more difficult to detect and contest because they operate under a powerful appearance of scientific objectivity. Algorithms are commonly perceived as neutral, data-driven, and impartial, a perception that can obscure the normative assumptions and structural inequalities embedded within their design and deployment. Such framing enables what critics describe as a form of tech-washing, in which discriminatory practices are legitimized through the ostensibly neutral language of data, models, and optimization \citep{Benjamin2019,Le2021}.

\subsection{Sources of Algorithmic Bias: Data, Design, and Power}\label{subsec1}

The sources of algorithmic bias are complex and interrelated, but they are often analytically distinguished as data-driven bias and design-driven bias \citep{Belenguer2022,Ferrara2023}. Data-driven bias emerges because machine-learning systems are trained on historical datasets that encode existing social inequalities, discriminatory practices, and institutionalized exclusions. As \citet{Chen2023} and \citet{Le2021} emphasize, training data are not neutral representations of reality. Instead, they are social artifacts shaped by earlier decisions about what is measured, how categories are defined, and whose experiences are made visible. Empirical research conducted by scholars at USC Viterbi illustrates the scale of this issue, showing that as many as 38.6 percent of widely used so-called common-sense facts incorporated into AI training data contain biased assumptions \citep{Gruel2022}.

These biases are not limited to individual data points but extend to the classificatory frameworks that organize datasets. Binary gender classifications exclude non-binary and gender-diverse identities, while racially coded language and proxy variables reproduce long-standing stereotypes and inequalities \citep{Buolamwini2018,Varsha2023}. Data collection and labeling are also inherently social and political processes. Decisions concerning ground truth, annotation standards, and population coverage are shaped by institutional priorities, economic constraints, and the positionality of data workers, often reinforcing existing power asymmetries \citep{Eubanks2018,Hanna2025}.

Design-driven bias arises from decisions made by system designers, developers, and organizations, including choices about feature selection, optimization objectives, and evaluation metrics. A growing body of scholarship links such biases to the lack of diversity within the AI workforce, suggesting that relatively homogeneous design teams are more likely to overlook harms experienced by marginalized groups \citep{Gardner2022,Hanna2025}. These risks are further intensified by commercial incentives that prioritize efficiency, scalability, and profitability over fairness, accountability, and social responsibility \citep{ONeil2016,Baker2022}. Together, biased data and biased design reinforce one another, producing systems that are both inequitable and difficult to scrutinize.

Opacity itself plays a central role in the persistence of algorithmic bias. Many AI systems operate as black boxes, which limits the ability of affected individuals, regulators, and even system developers to understand or contest their outputs. This lack of transparency undermines accountability, particularly in high-stakes settings where algorithmic decisions have lasting consequences for people's lives \citep{Ulenaers2020,Karnouskos2021}.

\subsection{Documented Human Harms Across Domains}\label{subsec2}

The societal impacts of algorithmic bias are well documented across a range of sectors. In criminal justice, ProPublica's investigation of the COMPAS risk-assessment tool revealed systematic racial disparities. The system disproportionately misclassified Black defendants as high risk while underestimating risk among white defendants \citep{Angwin2016}. These errors translate into longer sentences, harsher bail conditions, and the reinforcement of racial inequalities. Such practices raise serious concerns about due process and the right to a fair trial, particularly when proprietary algorithms cannot be meaningfully examined or challenged \citep{Ulenaers2020,Johansson2011}.

In employment contexts, algorithmic bias has been shown to reproduce and intensify gender inequality. Amazon's now-abandoned AI recruiting tool learned to disadvantage women by penalizing gender-coded terms and downgrading applicants associated with women's colleges after being trained on historical hiring data drawn from a male-dominated industry \citep{Dastin2018}. Similar dynamics have been identified across automated hiring and screening systems, prompting broader concerns about transparency and fairness in labor markets \citep{Andrews2022}.

\subsection{Theoretical Foundations of Algorithmic Bias}\label{subsec3}

The study of algorithmic bias is also a rapidly developing theoretical field. Scholars have developed formal definitions of fairness and analyzed the trade-offs between different fairness criteria. \citet{Wang2024} distinguishes multiple forms of algorithmic discrimination and examines the limits of regulatory responses in addressing harms that are structurally embedded. \citet{Chen2023} further refines this analysis through concepts such as predictive bias and agentic discrimination, emphasizing how AI systems can actively shape social inequalities rather than merely reflecting them.

These theoretical insights help explain why many technical remedies have proven insufficient. Approaches such as fairness through unawareness, which attempt to ensure fairness by excluding sensitive attributes like race or gender, have been widely criticized because algorithms can often infer these attributes from correlated variables \citep{Patty2022,Varsha2023}. More advanced fairness metrics, including demographic parity and equalized odds, seek to equalize outcomes or error rates across groups. However, these metrics frequently conflict with one another and with accuracy objectives. Impossibility results in algorithmic fairness demonstrate that it is mathematically impossible to satisfy all commonly desired fairness criteria at the same time, highlighting the inevitability of normative trade-offs \citep{Patty2022}.

These trade-offs reinforce a central conclusion: algorithmic bias cannot be resolved through technical solutions alone. Addressing it requires normative judgment, institutional accountability, and political decision-making about whose values and interests are prioritized. Algorithmic bias therefore exemplifies the algorithmic blind spot in practice, illustrating how ethical discourse becomes misaligned with lived human experience when speculative futures overshadow present harms. A genuinely human-centered approach to AI ethics must begin by confronting these documented injustices directly, rather than deferring them in favor of more futuristic debates about the moral status of machines.

This misalignment becomes especially visible in the unequal distribution of harm associated with biased AI systems. Algorithmic harms disproportionately affect racial minorities, women, and economically vulnerable populations, frequently intensifying existing inequalities \citep{Benjamin2019,Wang2024}. Rather than creating entirely new forms of injustice, algorithmic systems often reproduce historical patterns of exclusion while rendering them less visible and more difficult to contest. The perceived neutrality of algorithmic decision-making lends these outcomes an aura of inevitability, a dynamic that \citet{Benjamin2019} describes as the New Jim Code.

These patterns carry serious political implications. As algorithmic systems increasingly mediate access to employment, credit, healthcare, and legal outcomes, their biases erode trust in public institutions and deepen experiences of powerlessness among those most affected. The result extends beyond individual harm to include broader threats to democratic legitimacy. In this context, the algorithmic blind spot reflects a failure to recognize where ethical intervention is most urgently required.

\section{The Algorithmic Blind Spot in Focus: A Critical Synthesis}\label{sec4}

The preceding analysis of algorithmic harm provides the empirical basis for evaluating how ethical attention is distributed between speculative debates about artificial moral status and documented injustices. This comparative perspective makes it possible to examine whether patterns of discourse correspond to patterns of institutional uptake and measurable harm. The concurrent rise of robot rights discourse and the accumulation of empirical evidence documenting algorithmic bias is not accidental.

The figure of the artificial moral patient, imagined as a future entity capable of suffering and deserving moral or legal recognition, has proven philosophically compelling. However, this attraction carries significant ethical costs. When ethical inquiry is centered on hypothetical future harms to machines, urgent and ongoing injustices generated by biased, opaque, and unaccountable algorithmic systems risk being marginalized. 

This displacement of attention does not occur spontaneously; it is reinforced by identifiable cultural and institutional dynamics.
The redirection of ethical concern toward speculative AI futures is sustained by several mutually reinforcing dynamics. One influential factor is the cultural imagination shaped by science fiction, which has long portrayed AI through narratives involving sentient humanoid robots, existential threats, and profound moral disruption. These narratives render robot rights debates both familiar and emotionally resonant, drawing on enduring anxieties about humanity, autonomy, and technological transcendence \citep{Cave2019}. At the same time, philosophical inquiry often gravitates toward abstract questions of consciousness, personhood, and moral standing, which are intellectually engaging but frequently disconnected from institutional practice and social impact.

This orientation has tangible consequences. \citep{Gardner2022} and \citet{Baker2022} show that AI research funding structures tend to favor highly visible and future-oriented projects rather than work focused on bias mitigation, accountability, and governance. A feedback loop emerges in which speculative ethics attracts attention and resources, reinforcing the framing of AI as a potential moral subject, while research addressing harms to human populations remains comparatively underfunded. Media dynamics further amplify this imbalance, as narratives involving sentient AI or technological catastrophe are more likely to capture public attention than reporting on discriminatory credit systems, biased hiring tools, or automated welfare surveillance.

Commercial incentives also contribute to this pattern. Technology firms often find it more profitable to promote aspirational visions of advanced and human-like AI than to invest in the costly and politically complex work of addressing bias and institutional responsibility. The algorithmic blind spot therefore arises not from individual oversight but from structural conditions shaped by cultural fascination, disciplinary reward systems, media economies, and commercial interests.

\subsection{Operational Criteria}\label{subsec_operational}\label{subsec4a}

To render the algorithmic blind spot empirically assessable, we operationalize it through measurable cross-domain comparisons between two thematic corpora: speculative artificial moral-status research and empirically grounded bias-mitigation research. These corpora are constructed through bibliometric analysis of publication datasets extracted from large-scale bibliographic databases, enabling systematic comparison across publication volume, funding metadata, and policy-document linkage. The first comparison concerns discursive distribution: the overall volume of scholarly publications devoted to artificial moral status is comparable to, or exceeds, that devoted to documented algorithmic bias, indicating discursive prominence. The second comparison concerns funding density asymmetry: bias-mitigation research exhibits substantially higher grant density (grants per publication) than moral-status research, signaling stronger institutional investment in harm mitigation relative to discursive attention \citep{Whittaker2018,Gardner2022}. The third comparison concerns policy integration asymmetry: scholarship on algorithmic bias demonstrates significantly greater incorporation into policy documents and governance frameworks than scholarship on artificial moral status, reflecting differential translation into regulatory discourse \citep{UNESCO2021,ICRC2024}. Finally, these asymmetries must display temporal persistence across multiple years rather than appearing as short-term fluctuations. When these comparisons consistently reveal discursive parity alongside funding and policy integration asymmetries that persist over time, the blind spot becomes institutionally observable as a patterned misalignment between ethical discourse, resource allocation, and governance translation.

The following analysis applies these operational comparisons to bibliometric data in order to assess whether the proposed blind spot is empirically observable.

\subsection{Bibliometric Findings}\label{subsec4b}

Bibliometric analysis of the two thematic corpora yields the following results. The robot-rights / moral-status corpus contains 9,694 publications, associated with 1,591 total grants (1,384 unique grants) and 136 policy documents. The bias-mitigation corpus contains 8,769 publications, associated with 4,467 total grants (3,808 unique grants) and 902 policy documents. 

Publication volume is therefore comparable across the two domains (9,694 vs.\ 8,769 publications), yet substantial asymmetries emerge in institutional translation. The robot-rights corpus exhibits a grant density of approximately 0.143 unique grants per publication and a policy-integration ratio of approximately 0.014 policy documents per publication. By contrast, the bias-mitigation corpus exhibits a grant density of approximately 0.434 and a policy-integration ratio of approximately 0.103. Bias-mitigation research thus receives roughly three times the grant density and more than seven times the policy integration relative to publication volume. 

Year-by-year funding-intensity analysis confirms the persistence of this asymmetry across most of the observation window. Although the two curves partially converge in the final year, convergence reflects a decline from post-peak funding intensity in bias-mitigation research rather than a sustained equalization of grant density. Across the majority of the period examined, funding intensity remains structurally higher for bias-mitigation scholarship.

Taken together, the disparities point not to a difference in discursive presence, but to a structural asymmetry in institutional uptake and governance translation. Directional stability under alternative normalizations, including total versus unique grants and restricted temporal windows, further indicates that the asymmetry is not an artifact of metric specification.

\begin{figure}[H]
\centering
\includegraphics[width=0.85\linewidth]{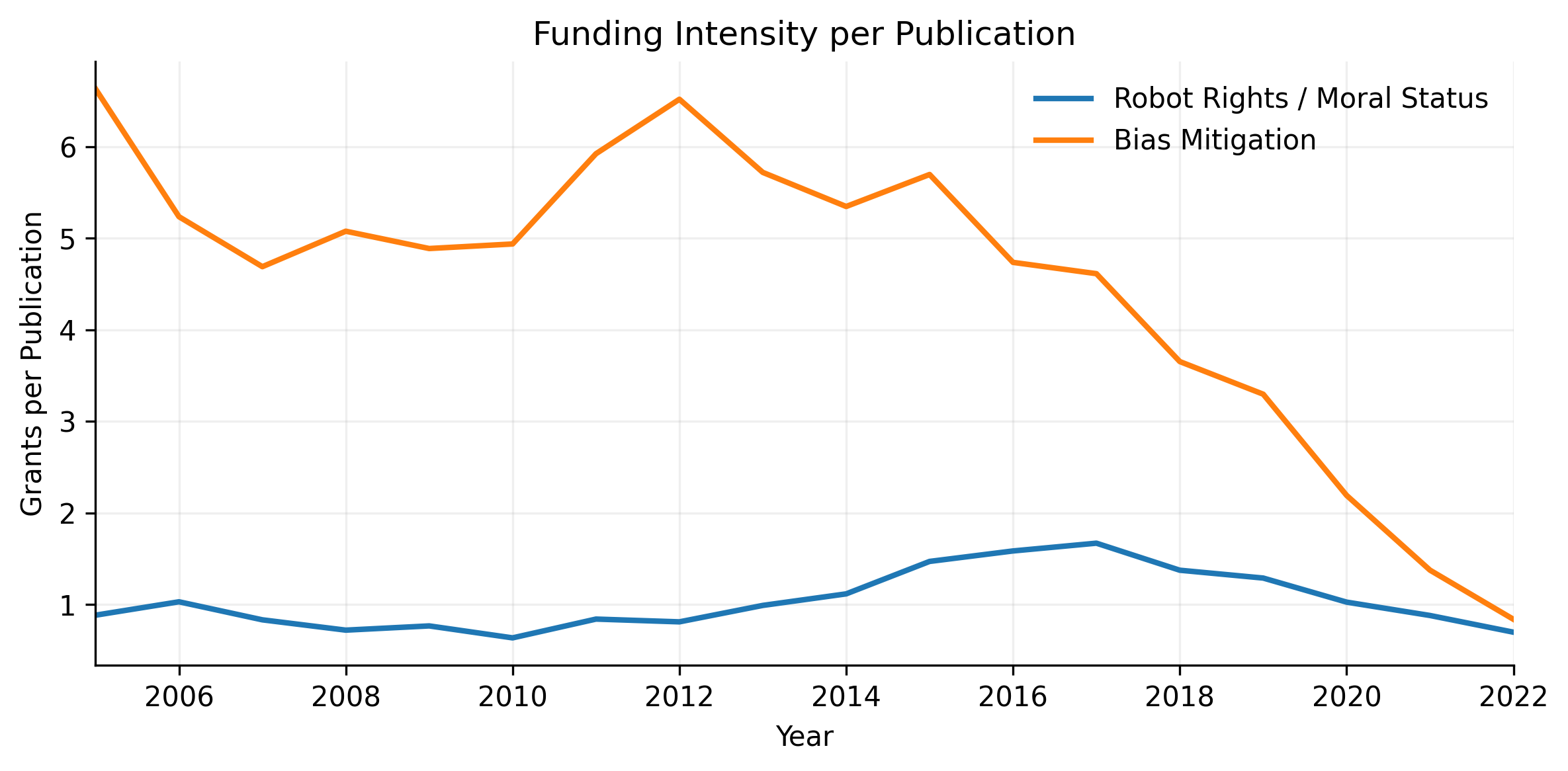}
\caption{Annual funding intensity (grants per publication) in robot-rights/moral-status and bias-mitigation research.}
\label{fig:funding_intensity}
\end{figure}

A parallel pattern emerges in policy integration intensity. When policy documents are normalized by publication volume on a year-by-year basis, bias-mitigation research exhibits consistently higher policy-integration density across the observation window. Short-term fluctuations occur, yet the gap remains structurally pronounced. The divergence is not confined to absolute policy-document counts; it reflects a persistent disparity in how discursive production translates into governance uptake. The blind spot therefore appears as an asymmetry in institutional integration rather than a difference in scholarly activity per se.

\begin{figure}[H]
\centering
\includegraphics[width=0.85\linewidth]{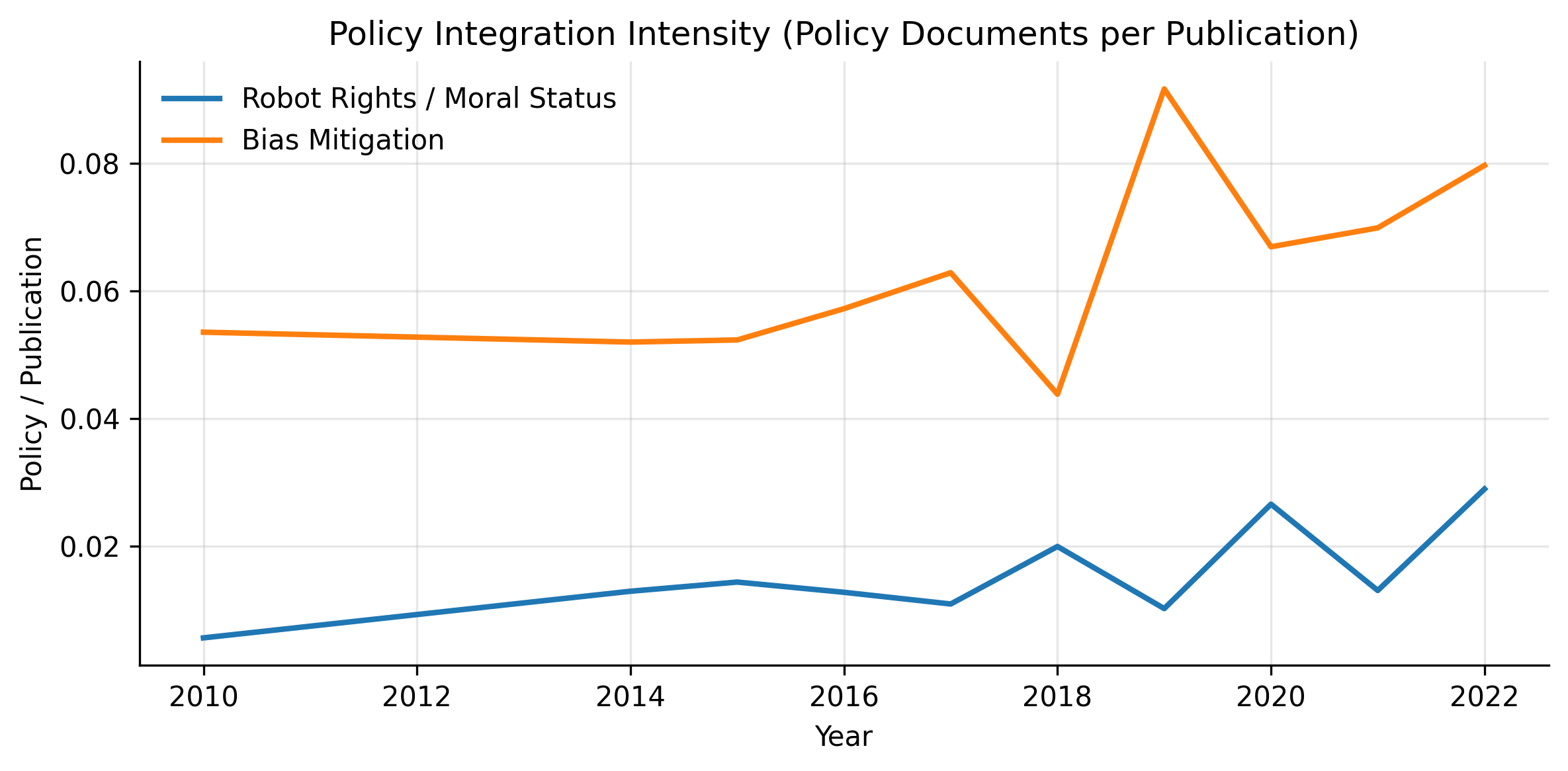}
\caption{Annual policy-integration intensity (policy documents per publication) in robot-rights/moral-status research and bias-mitigation research.}
\label{fig:policy_intensity}
\end{figure}

Empirical evidence thus grounds the operational comparisons introduced above. Discursive production across the two domains remains comparable, whereas institutional translation diverges substantially. The blind spot does not consist in the absence of funding or policy engagement with bias-mitigation research; rather, it manifests in the coexistence of discursive parity with pronounced asymmetries in grant density and policy integration. The concept is therefore empirically testable and institutionally observable, rather than merely interpretive.

Temporal persistence was assessed by examining publication, grant, and policy-linkage trajectories across the observation window. Relative stability of publication parity alongside sustained asymmetries in grant density and policy integration indicates a durable structural configuration rather than a short-term fluctuation. Persistence over time reinforces interpretation of the blind spot as a patterned misalignment rather than a contingent anomaly.

\begin{figure}[H]
\centering
\includegraphics[width=0.9\linewidth]{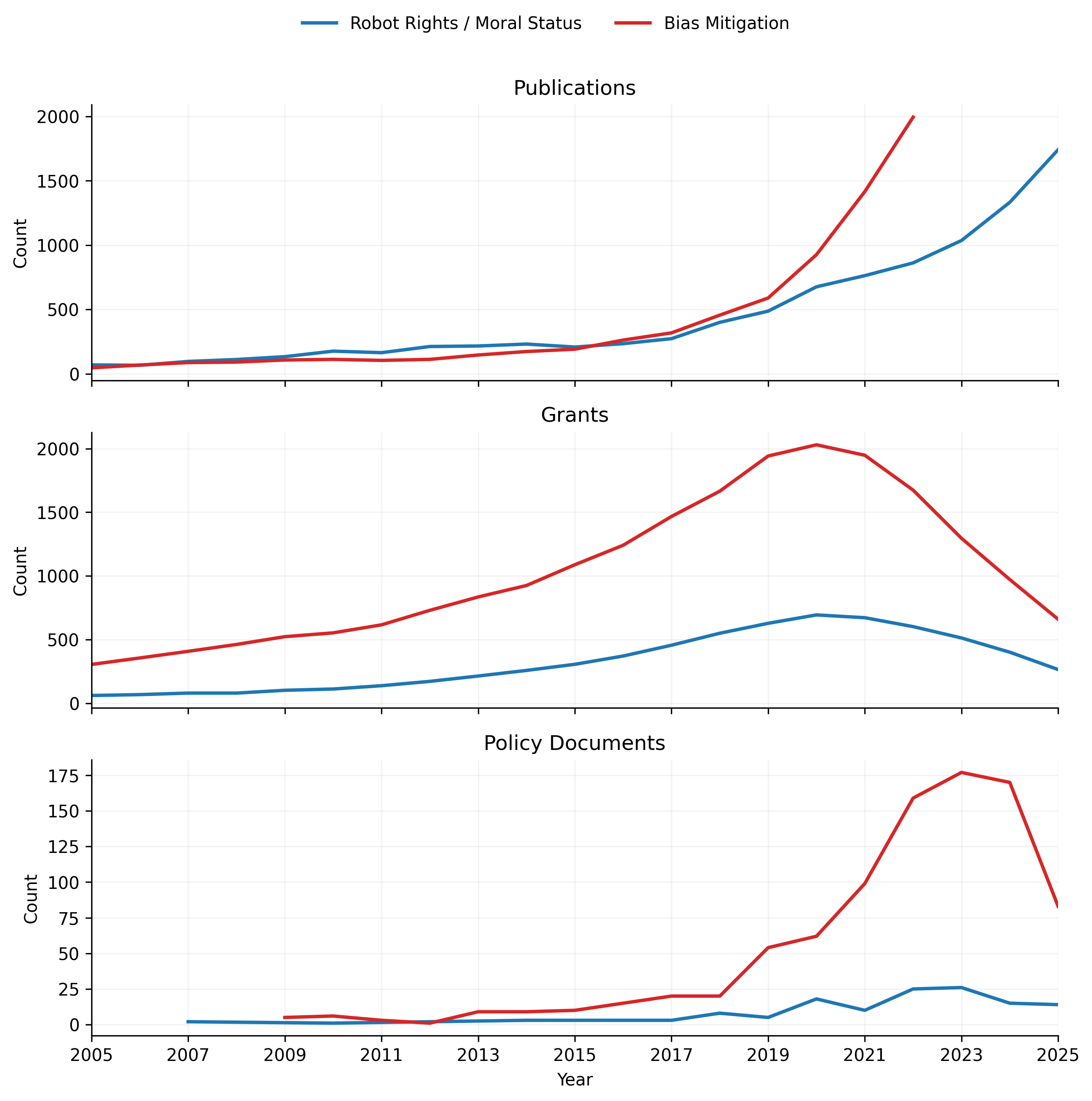}
\caption{Year-by-year comparison of publications, grants, and policy documents in robot-rights/moral-status research and bias-mitigation research.}
\label{fig:yearly_comparison}
\end{figure}

Empirical results should not be read as implying that attention to speculative questions necessarily displaces present-focused ethical analysis, as in a ``zero-sum'' framing, nor that speculative reflection on artificial agency lacks philosophical legitimacy. The analysis does not argue against robot-rights scholarship as such, nor does it deny the value of anticipatory governance in conditions of technological uncertainty. Rather, the evidence identifies a specific institutional pattern: discursive parity between domains can coexist with pronounced asymmetries in funding density and policy integration. The concept of the algorithmic blind spot does not reject speculative ethics; it isolates a structural misalignment between discursive production and governance translation that becomes visible when publication volume is examined alongside institutional uptake.

\subsection{Interpretation and Counterarguments}\label{subsec4c}

The empirical analysis does not demonstrate that attention to speculative ethics mechanically displaces research on algorithmic harm. It shows a different pattern: comparable levels of publication activity coexist with substantial differences in funding density and policy integration. The claim concerns patterns of institutional translation rather than individual intent. It concerns how ethical discourse translates into governance structures.

Speculative inquiry into artificial moral status remains philosophically legitimate. Questions about future machine agency and consciousness are intellectually serious and may acquire practical relevance. The argument advanced here does not deny their value. 
It asks whether institutional priorities correspond to where documented harms are actually occurring.

Algorithmic systems already influence sentencing decisions, employment screening, credit allocation, and public surveillance. The harms associated with these systems are measurable and disproportionately borne by vulnerable populations. In such conditions, proportional alignment between discursive production and institutional embedding becomes ethically significant.

The diagnosis of an algorithmic blind spot therefore identifies a structural imbalance in governance translation. It does not portray robot-rights scholarship as misguided, nor does it deny the relevance of anticipatory governance. It distinguishes philosophical exploration from institutional prioritization. The central concern is sequencing and proportionality under conditions of demonstrable harm.

The cases analyzed in Section~\ref{sec3} provide the empirical backdrop for this claim. They illustrate that the harms in question are measurable, institutionally embedded, and socially distributed. The contrast with speculative harms is therefore not rhetorical but temporal and structural.

In this context, appeals to precaution in favor of artificial moral patients must also confront the asymmetry of certainty involved. As \citet{Zuradzki2021} notes, extending precaution to entities whose moral status is indeterminate raises difficult justificatory questions. By contrast, the harms produced by biased systems are empirically documented. A coherent application of precaution therefore requires attention to risks that are already measurable and socially distributed.

\section{Reorienting AI Ethics: A Human-Centric Framework for a Just Future}\label{sec5}

The preceding analysis has identified a structural imbalance in ethical prioritization. The task now is to translate that diagnosis into institutional and normative guidance.

Addressing the algorithmic blind spot requires a fundamental reorientation of AI ethics. Rather than privileging speculative, machine-centered concerns, ethical inquiry must be grounded in the protection of human welfare. This does not entail abandoning philosophical questions about AI consciousness, but subordinating them to the urgent need to ensure that AI systems are fair, accountable, and socially beneficial. This section proposes a human-centric framework organized around three interdependent pillars: fairness by design, transparency and explainability, and accountability with effective mechanisms for redress.

A reoriented AI ethics must begin from the primacy of human welfare. As \citet{Birhane2020} argue, when human well-being is treated as central, the ethical failures of AI design and deployment become unavoidable. A human-centric approach recognizes that any consideration of artificial moral status must remain secondary to the protection of human rights. As \citet{DeStefano2019} shows in the context of labor automation, abstract ethical principles offer little protection when material rights are at stake. Ethical evaluation should therefore prioritize the concrete effects of AI systems on human lives.

The first pillar of this framework is \emph{fairness by design} and proactive bias mitigation. Rather than responding to harms after deployment, ethical governance should embed fairness into system development from the outset \citep{Ferrara2023}. This includes diversifying training data, applying debiasing techniques, and incorporating interdisciplinary expertise \citep{Zewe2024,Soleimani2025}. Achieving this goal requires a cultural shift within AI development toward prioritizing equity and responsibility.

The second pillar is \emph{transparency and meaningful explainability}. The opacity of many AI systems undermines accountability and public trust. Transparency is not merely technical but ethical, enabling affected individuals to understand and contest algorithmic decisions \citep{Chen2023,Belenguer2022}. \citet{Schwitzgebel2018} further argues that AI systems must not misrepresent their capabilities or moral status, as doing so risks deception and harm.

The third pillar is \emph{accountability and redress}. Clear responsibility must exist when AI systems cause harm, supported by legal frameworks, oversight mechanisms, and accessible remedies \citep{Birhane2020}. The COMPAS case illustrates the consequences of accountability failure \citep{Angwin2016}. Addressing systemic harm may require new legal tools, including rights to explanation and institutional audits.

These pillars are mutually reinforcing and must be pursued together. Technical solutions alone are insufficient. A human-centric AI ethics also requires coordinated action by policymakers, funding agencies, and international organizations. Regulatory efforts such as the EU AI Act exemplify a risk-based, human-focused approach, while organizations like \citet{UNESCO2021} and \citet{ICRC2024} provide normative guidance. Aligning research incentives with human-centric goals and resisting narratives that prioritize disruption over justice are essential steps toward an AI ethics that serves humanity rather than harms it.

\subsection{Institutional Implications}\label{subsec5a}

One implication of the foregoing analysis concerns funding structures. When publication levels are similar but funding and policy uptake differ substantially, funding agencies should increase transparency about how resources are allocated across themes. Periodic review of grant density across major AI ethics domains could help ensure that institutional support remains responsive to documented social impact rather than solely to emerging conceptual trends.

A second implication concerns policy advisory design. When regulatory frameworks are developed, balanced representation across domains of AI ethics becomes important. Incorporating expertise on algorithmic harm, bias mitigation, and governance alongside speculative ethics may reduce the risk that forward-looking debates overshadow present institutional responsibilities.

A further implication involves ongoing bibliometric monitoring. Systematic tracking of publication volume, funding density, and policy citation patterns can provide an empirical basis for assessing whether governance translation aligns proportionately with documented harms. Such monitoring does not dictate outcomes but introduces institutional reflexivity.

Finally, the diagnosis supports the expansion of harm-audit frameworks for deployed systems. Where algorithmic tools materially affect sentencing, employment, credit, or access to services, structured review mechanisms and accountability procedures become central to aligning ethical discourse with operational practice.

\section{Conclusion: Overcoming the Algorithmic Blind Spot}\label{sec6}

This paper has argued that contemporary AI ethics exhibits a structural imbalance in moral attention. Debates concerning the future moral status of artificial intelligence coexist with empirically documented harms produced by biased and unaccountable algorithmic systems already in operation. The algorithmic blind spot names the institutional misalignment that arises when discursive production does not correspond to patterns of governance integration and measurable harm.

The imbalance identified here is sustained at the level of discourse, funding, and policy translation. While speculative debates focus on uncertain future harms, algorithmic systems currently shape access to employment, credit, legal outcomes, and public services \citep{ONeil2016,Eubanks2018,Benjamin2019}. These harms are socially patterned and institutionally embedded. When normative attention diverges from documented impact, ethical reflection risks losing alignment with lived consequences.

dentifying the blind spot does not require abandoning philosophical inquiry into artificial moral status. It requires clarifying the relationship between speculative ethics and institutional responsibility. When anticipatory debates expand without corresponding attention to present harms, ethical prioritization becomes distorted. From an STS perspective, this distortion reflects a disconnect between ethical imagination and the socio-technical systems through which algorithmic power operates \citep{Jasanoff2004,Latour2005}.

Reorienting AI ethics requires sustained institutional alignment with documented harms. Technical mitigation alone is insufficient; effective governance depends on normative judgment, oversight mechanisms, and enforceable accountability structures \citep{Whittaker2018,Gardner2022}. Regulatory initiatives such as the European Union’s risk-based model and international frameworks advanced by UNESCO and the ICRC reflect movement toward human-centered governance \citep{UNESCO2021,ICRC2024}.

Under conditions of uncertainty and high stakes, ethical reasoning must remain anchored to demonstrable risk and institutional impact \citep{Funtowicz1993}. The expansion of speculative moral concern should not eclipse accountability for systems already shaping social outcomes. The central question for AI ethics is therefore not whether artificial agents may one day merit rights, but whether existing algorithmic systems are governed in ways that uphold human dignity, equality, and democratic legitimacy.

The broader implication of this analysis is that the algorithmic blind spot---understood as the systematic reorientation of ethical attention away from present, empirically substantiated harms to human populations and toward speculative future artificial moral subjects---cannot be addressed through an expansion of the moral circle alone. Ethical progress in the domain of artificial intelligence requires, at minimum, sustained attention to ensuring that those already encompassed within that circle are not rendered invisible through processes of abstraction, automation, or institutional neglect. Absent demonstrable commitments by algorithmic systems to uphold human dignity, substantive equality, and democratic accountability, debates concerning robot rights risk operating as a form of moral displacement, regardless of their philosophical sophistication.

Addressing the algorithmic blind spot is thus a necessary condition for the development of a responsible and credible AI ethics. Doing so requires a reorientation of scholarly priorities toward sustained empirical investigation of algorithmic bias, governance, and institutional impact; a corresponding commitment by system designers to incorporate principles of fairness, transparency, and contestability into technical architectures; and robust policy interventions capable of ensuring meaningful accountability and avenues for redress \citep{Whittaker2018,Gardner2022,UNESCO2021}. From a Science and Technology Studies perspective, this challenge is neither purely technical nor solely moral, but fundamentally infrastructural: ethical reflection must remain closely entangled with the socio-technical systems and institutional arrangements through which algorithmic power is produced and exercised. Framed in this way, the central ethical question raised by artificial intelligence is not whether machines may one day warrant moral or legal rights, but whether human societies can effectively govern the systems they have already created. The resolution of this question will play a decisive role in determining whether AI functions as an instrument of justice or as a mechanism for the reproduction and normalization of inequality.

\bibliography{references}

@article{Andrews2022,
  author = {Andrews, L. and Bucher, H.},
  title = {Automating discrimination: {AI} hiring practices and gender inequality},
  journal = {Cardozo Law Review},
  volume = {44},
  number = {1},
  pages = {145--202},
  year = {2022}
}

@misc{Angwin2016,
  author = {Angwin, J. and Larson, J. and Mattu, S. and Kirchner, L.},
  title = {Machine Bias: Risk Assessments in Criminal Sentencing},
  howpublished = {ProPublica},
  year    = {2016},
  url     = {https://www.propublica.org/article/machine-bias-risk-assessments-in-criminal-sentencing},
  note    = {Accessed: 2026-03-02}
}

@misc{Baker2022,
  author = {Baker, D. and Hanna, A.},
  title = {{AI} Ethics Are in Danger. {F}unding Independent Research Could Help},
  howpublished = {Stanford Social Innovation Review},
  year = {2022},
  url     = {https://ssir.org/articles/entry/ai_ethics_are_in_danger_funding_independent_research_could_help},
  note    = {Accessed: 2026-03-02}
}

@article{Belenguer2022,
  author  = {Belenguer, L.},
  title   = {{AI} bias: {E}xploring discriminatory algorithmic decision-making models and the application of the non-discrimination principle within a human rights framework},
  journal = {AI Ethics},
  year    = {2022},
  volume  = {2},
  number  = {4},
  pages   = {771--787},
  doi     = {10.1007/s43681-022-00138-8}
}

@book{Benjamin2019,
  author    = {Benjamin, R.},
  title     = {Race After Technology},
  year      = {2019},
  publisher = {Polity},
  address   = {Cambridge}
}

@inproceedings{Birhane2020,
  author    = {Birhane, A. and van Dijk, J.},
  title     = {Robot Rights? {L}et's Talk about Human Welfare Instead},
  booktitle = {AIES '20: Proceedings of the AAAI/ACM Conference on AI, Ethics, and Society},
  year      = {2020},
  pages     = {207--213},
  doi       = {10.1145/3375627.3375855}
}

@article{Bryson2018,
  author = {Bryson, J. J.},
  title = {Patiency is not a virtue: {T}he design of intelligent systems and systems of ethics},
  journal = {Ethics and Information Technology},
  year = {2018},
  volume  = {20},
  number  = {1},
  pages   = {15--26},
  doi     = {10.1007/s10676-018-9448-6}
}

@inproceedings{Buolamwini2018,
  author    = {Buolamwini, J. and Gebru, T.},
  title     = {Gender Shades: Intersectional Accuracy Disparities in Commercial Gender Classification},
  booktitle = {Proceedings of the 1st Conference on Fairness, Accountability and Transparency},
  series    = {Proceedings of Machine Learning Research},
  volume    = {81},
  pages     = {77--91},
  year      = {2018},
  url       = {https://proceedings.mlr.press/v81/buolamwini18a.html}
}

@inproceedings{Cave2019,
  author = {Cave, S. and Dihal, K. and Dillon, S.},
  title = {{AI} narratives},
  booktitle = {Proceedings of the AAAI/ACM Conference on AI, Ethics, and Society},
  year = {2019},
  pages     = {387--392},
  doi       = {10.1145/3306618.3314231}
}

@article{Chen2023,
  author  = {Chen, Z.},
  title   = {Ethics and discrimination in artificial intelligence-enabled recruitment practices},
  journal = {Humanities and Social Sciences Communications},
  year    = {2023},
  volume  = {10},
  number  = {1},
  pages   = {567},
  doi     = {10.1057/s41599-023-02079-x}
}

@article{Danaher2020,
  author  = {Danaher, J.},
  title   = {Welcoming Robots into the Moral Circle: {A} Defence of Ethical Behaviourism},
  journal = {Science and Engineering Ethics},
  year    = {2020},
  volume  = {26},
  number  = {4},
  pages   = {2023--2049},
  doi     = {10.1007/s11948-019-00119-x}
}

@misc{Dastin2018,
  author = {Dastin, J.},
  title = {Amazon scraps secret {AI} recruiting tool that showed bias against women},
  howpublished = {Reuters},
  year = {2018},
  url = {https://www.reuters.com/article/us-amazon-com-jobs-automation-insight-idUSKCN1MK08G},
  note = {Accessed: 2026-03-02}
}

@article{DeStefano2019,
  author  = {De Stefano, V.},
  title   = {``{N}egotiating the {A}lgorithm'': {A}utomation, Artificial Intelligence, and Labor Protection},
  journal = {Comparative Labor Law \& Policy Journal},
  year    = {2019},
  volume  = {41},
  number  = {1},
  pages   = {1--32}
}

@book{Eubanks2018,
  author    = {Eubanks, V.},
  title     = {Automating Inequality: How High-Tech Tools Profile, Police, and Punish the Poor},
  year      = {2018},
  publisher = {St.\ Martin's Press},
  address   = {New York}
}

@article{Ferrara2023,
  author  = {Ferrara, E.},
  title   = {Fairness and Bias in Artificial Intelligence: {A} Brief Survey of Sources, Impacts, and Mitigation Strategies},
  journal = {Sci},
  year    = {2023},
  volume  = {6},
  number  = {1},
  pages   = {3},
  doi     = {10.3390/sci6010003}
}

@book{Floridi2013,
  author    = {Floridi, L.},
  title     = {The Ethics of Information},
  year      = {2013},
  publisher = {Oxford University Press},
  address   = {Oxford},
  doi       = {10.1093/acprof:oso/9780199641321.001.0001}
}

@book{Foucault1977,
  author = {Foucault, M.},
  title = {Discipline and Punish: The Birth of the Prison},
  publisher = {Pantheon Books},
  address   = {New York},
  year      = {1977}
}

@book{Foucault1980,
  author = {Foucault, M.},
  title = {Power/Knowledge: Selected Interviews and Other Writings 1972-1977},
  publisher = {Pantheon Books},
  address   = {New York},
  year      = {1980}
}

@article{Funtowicz1993,
  author = {Funtowicz, S. O. and Ravetz, J. R.},
  title = {Science for the post-normal age},
  journal = {Futures},
  year = {1993},
  volume = {25},
  number = {7},
  pages = {739--755},
  doi = {10.1016/0016-3287(93)90022-L}
}

@article{Gardner2022,
  author  = {Gardner, A. and Smith, A. L. and Steventon, A. and Coughlan, E. and Oldfield, M.},
  title   = {Ethical funding for trustworthy {AI}: proposals to address the responsibilities of funders to ensure that projects adhere to trustworthy {AI} practice},
  journal = {AI and Ethics},
  year    = {2022},
  volume  = {2},
  pages   = {277--291},
  doi     = {10.1007/s43681-021-00069-w}
}

@misc{Gruel2022,
  author = {Gruel, M.},
  title = {That's just common sense: {USC} researchers find bias in up to 38.6% of facts used by AI},
  howpublished = {USC Viterbi School of Engineering},
  year = {2022},
  url     = {https://viterbischool.usc.edu/news/2022/05/thats-just-common-sense-usc-researchers-find-bias-in-up-to-38-6-of-facts-used-by-ai/},
  note    = {Accessed: 2026-03-02}
}

@article{Hanna2025,
  author  = {Hanna, M. G.},
  title   = {Ethical and Bias Considerations in Artificial Intelligence/Machine Learning},
  journal = {Modern Pathology},
  year    = {2025},
  volume  = {38},
  number  = {3},
  pages   = {100686},
  doi     = {10.1016/j.modpat.2024.100686}
}

@misc{ICRC2024,
  author = {{ICRC}},
  title = {The problem of algorithmic bias and military applications of {AI}},
  year = {2024},
  url     = {https://blogs.icrc.org/law-and-policy/2024/03/14/falling-under-the-radar-the-problem-of-algorithmic-bias-and-military-applications-of-ai/},
  note    = {Accessed: 2026-03-02}
}

@book{Jasanoff2004,
  author    = {Jasanoff, S.},
  title     = {States of Knowledge: The Co-Production of Science and Social Order},
  year      = {2004},
  publisher = {Routledge},
  address   = {London}
}

@article{Johansson2011,
  author  = {Johansson, L.},
  title   = {Is It Morally Right to Use Unmanned Aerial Vehicles ({UAV}s) in War?},
  journal = {Philosophy \& Technology},
  year    = {2011},
  volume  = {24},
  number  = {3},
  pages   = {279--291},
  doi     = {10.1007/s13347-011-0033-8}
}

@article{Karnouskos2021,
  author  = {Karnouskos, S.},
  title   = {Symbiosis with artificial intelligence via the prism of law, robots, and society},
  journal = {Artificial Intelligence and Law},
  year    = {2021},
  volume  = {30},
  number  = {1},
  pages   = {93--115},
  doi     = {10.1007/s10506-021-09289-1}
}

@book{Latour1993,
  author = {Latour, B.},
  title = {We Have Never Been Modern},
  publisher = {Harvard University Press},
  address   = {Cambridge, MA},
  year      = {1993}
}

@book{Latour2005,
  author = {Latour, B.},
  title = {Reassembling the Social: An Introduction to Actor-Network-Theory},
  publisher = {Oxford University Press},
  address   = {Oxford},
  year      = {2005}
}

@misc{Le2021,
  author = {Le, V.},
  title = {Algorithmic Bias Explained: How Automated Decision-Making Becomes Automated Discrimination},
  howpublished = {Greenlining Institute},
  year = {2021},
  url     = {https://greenlining.org/publications/algorithmic-bias-explained/},
  note    = {Accessed: 2026-03-02}
}

@book{ONeil2016,
  author    = {O'Neil, C.},
  title     = {Weapons of Math Destruction},
  year      = {2016},
  publisher = {Crown},
  address   = {New York}
}

@article{Patty2022,
  author  = {Patty, J. W. and Penn, E. M.},
  title   = {Algorithmic Fairness and Statistical Discrimination},
  journal = {Philosophy Compass},
  year    = {2022},
  volume  = {18},
  number  = {1},
  pages   = {e12891},
  doi     = {10.1111/phc3.12891}
}

@article{Schwitzgebel2018,
  author = {Schwitzgebel, E.},
  title = {The moral status of artificial minds},
  journal = {Journal of Consciousness Studies},
  year = {2018},
  volume = {25},
  number = {9-10},
  pages = {172--185}
}

@article{Soleimani2025,
  author  = {Soleimani, M.},
  title   = {Reducing {AI} bias in recruitment and selection: {A} grounded theory approach},
  journal = {The International Journal of Human Resource Management},
  year    = {2025},
  volume  = {36},
  number  = {14},
  pages   = {2480--2515},
  doi     = {10.1080/09585192.2025.2480617}
}

@article{Turing1950,
  author = {Turing, A. M.},
  title = {Computing Machinery and Intelligence},
  journal = {Mind},
  year = {1950},
  volume = {LIX},
  number = {236},
  pages = {433--460},
  doi = {10.1093/mind/LIX.236.433}
}

@article{Ulenaers2020,
  author  = {Ulenaers, J.},
  title   = {The Impact of Artificial Intelligence on the Right to a Fair Trial: Towards a Robot Judge?},
  journal = {Asian Journal of Law and Economics},
  year    = {2020},
  volume  = {11},
  number  = {2},
  pages   = {20200008},
  doi     = {10.1515/ajle-2020-0008}
}

@techreport{UNESCO2021,
  author      = {{UNESCO}},
  title       = {Recommendation on the Ethics of Artificial Intelligence},
  institution = {United Nations Educational, Scientific and Cultural Organization},
  address     = {Paris},
  year        = {2021},
  url         = {https://www.unesco.org/en/artificial-intelligence/recommendation-ethics},
  note        = {Accessed: 2026-03-02}
}

@article{Varsha2023,
  author  = {Varsha, P. S.},
  title   = {How can we manage biases in artificial intelligence systems -- {A} systematic literature review},
  journal = {International Journal of Information Management Data Insights},
  year    = {2023},
  volume  = {3},
  number  = {1},
  pages   = {100165},
  doi     = {10.1016/j.jjimei.2023.100165}
}

@article{Wang2024,
  author  = {Wang, X. and Wu, Y. C. and Ji, X. and Fu, H.},
  title   = {Algorithmic discrimination: {E}xamining its types and regulatory measures with emphasis on {US} legal practices},
  journal = {Frontiers in Artificial Intelligence},
  year    = {2024},
  volume  = {7},
  pages   = {1320277},
  doi     = {10.3389/frai.2024.1320277}
}

@techreport{Whittaker2018,
  author      = {Whittaker, M. and Crawford, K. and Dobbe, R. and Fried, G. and Kaziunas, E. and Mathur, V. and {Myers West}, S. and Richardson, R. and Schultz, J. and Schwartz, O.},
  title       = {{AI} Now Report 2018},
  institution = {AI Now Institute, New York University},
  address     = {New York},
  year        = {2018},
  url         = {https://ainowinstitute.org/wp-content/uploads/2023/04/AI_Now_2018_Report.pdf}
}

@article{Wynter2003,
  author  = {Wynter, S.},
  title   = {Unsettling the Coloniality of {Being/Power/Truth/Freedom}: Towards the Human, After Man, Its Overrepresentation---{A}n Argument},
  journal = {CR: The New Centennial Review},
  year    = {2003},
  volume  = {3},
  number  = {3},
  pages   = {257--337},
  doi     = {10.1353/ncr.2004.0015}
}

@misc{Zewe2024,
  author = {Zewe, A.},
  title = {Researchers reduce bias in {AI} models while preserving or improving accuracy},
  howpublished = {MIT News},
  year = {2024},
  url     = {https://news.mit.edu/2024/researchers-reduce-bias-ai-models-while-preserving-improving-accuracy-1211},
  note    = {Accessed: 2026-03-02}
}

@article{Zuradzki2021,
  author  = {{\. Z}uradzki, T.},
  title   = {Against the Precautionary Approach to Moral Status},
  journal = {The American Journal of Bioethics},
  year    = {2021},
  volume  = {21},
  number  = {1},
  pages   = {53--56},
  doi     = {10.1080/15265161.2020.1845868}
}

\end{document}